\documentclass[12pt,a4paper]{article}%
\usepackage{amsmath,amssymb,amsfonts}
\usepackage{graphicx,epsfig,url}
\usepackage{color}
\usepackage{amsmath}
\usepackage{amsfonts}
\usepackage{amssymb}
\usepackage{float}
\usepackage{caption2}
\usepackage{graphicx}
\usepackage{hyperref}%
\numberwithin{equation}{section} 
\setcaptionmargin{1cm}

\begin{document}
\begin{titlepage}
\begin{flushright}
IFUP--TH/2009-2\hfill
\end{flushright}
\vskip 1.0cm
\begin{center}
{\Large \bf Multi-muon events at the Tevatron:\\[5mm] a hidden sector
from hadronic collisions} \vskip 1.0cm {\large Riccardo
Barbieri$^a$,\ \   Lawrence J.~Hall$^b$, \\[10pt]
Vyacheslav S.~Rychkov$^a$ and Alessandro Strumia$^c$} \\[1cm]
{\it $^a$ Scuola Normale Superiore and INFN, Piazza dei Cavalieri 7, 56126 Pisa, Italy} \\[5mm]
{\it $^b$ Department of Physics, University of California, Berkeley, and\\
Theoretical Physics Group, LBNL, Berkeley, CA 94720, USA}\\[5mm]
{\it $^c$ Dipartimento di Fisica dell'Universit\`a di Pisa and INFN,
Italia} \vskip 1.0cm
\end{center}
\begin{abstract}
We show an explicit attempt to interpret the multi-muon anomaly recently
 claimed by the CDF collaboration in terms of a light
scalar singlet $\phi$ which communicates with the standard quarks
either through a heavy scalar or a heavy fermion exchange.
Building on arXiv:0810.5730,  that suggested a singlet $\phi$ with a chain decay
into a final state made of four $\tau\bar{\tau}$ pairs,
we can simulate most of the muon properties of the selected sample of
events. Some of these properties adhere rather well to the already
published data;  others should allow a decisive test of the proposed
interpretation. Assuming that the test is positively passed, we show
how the PAMELA excess can be fitted by the annihilation of a TeV
Dark Matter particle that communicates with the Standard Model via
the new light singlet(s).
\end{abstract}
\end{titlepage}

\section{Introduction and statement of the problem}

The CDF Collaboration has recently published a study of a significant sample
of multi-muon events with unexpected properties \cite{Giro-big}.
The D0 Collaboration performed a similar search without finding similar events~\cite{D0}.
While it is
still possible that these events will be in the end accounted for in terms of
standard physics and detector effects, yet a small but significant fraction of
them has characteristics which are peculiar enough to deserve attention. This
has in fact prompted a phenomenological conjecture \cite{Giro-small} that
tries to explain them in terms of new physics: the pair production of a
relatively light Standard Model singlet, $\phi$, each with a cascade decay
into a final state made of 8 tau leptons. An intriguing feature of this
\textit{explanation} is that it describes well the sign-coded multiplicity
distribution of the additional muons, from two on, found in a
cone\footnote{The events in question are triggered by a pair of muons with
$p_{T}>3$ GeV and $|\eta|<0.7$. Additional muons with $p_{T}>2$ GeV and
$|\eta|<1.1$, which are often present in these events, are grouped into two
\textit{cones} of opening angle 36.8$^{\circ}$ around the directions of the
trigger (also called \textit{primary}) muons. Muons which happen to lie
outside of these cones are ignored.} around the direction of a primary muon.
In this interpretation, such a distribution is simply related, given the
detector efficiencies in muon identification, to the $\tau\rightarrow\mu$
decay probability. On the other hand, the apparent difficulty in exhibiting a
plausible production mechanism of the $\phi$-pair casts doubts on the
significance of the proposal and, even more importantly, prevents a decisive
comparison of the model with the data. In spite of its uncertainties and
ambiguities, this situation motivates us to try to go further.

\smallskip

A main problem one faces right at the beginning is how to select, from the
overwhelming number of background events, those ones that may represent the
signal. Secondly, a proper understanding of the event properties that involve
hadronic tracks requires a detection simulation that we cannot do. To get
around these problems we adopt the following strategy. We first concentrate
our attention on a sub-class of about 4000
events that contain at least two muons (i.e.\ one
primary and at least one additional) in each of the two cones. These events
have also been highlighted in \cite{Giro-big,Giro-small} and found to have
several properties especially difficult to understand in terms of known
physics. We assume that these events constitute \textit{the signal} and we
seek for a production mechanism of the $\phi$-pair that can account for the
measured properties of the muons contained in them. Out of all possibilities,
discussed in Section 2, somewhat surprisingly one single effective operator of
dimension 5 fits the invariant mass distribution of all muons in this sample
for a definite value of its scale. The operator is of the form $\bar{q}%
q\phi^{2}$, where $q$ is a first generation quark.

Encouraged by this result, which fixes the production mechanism, we analyze in
Section 3 a few other measured features of the multi-muon events: the $\mu^\pm$ multiplicity,
impact parameter and invariant mass of muons in one cone.
Given our
definition of the signal, the background can be clearly identified.
%While, as expected, the significant presence of background events is confirmed, some
%features of the model are strengthened and some others emerge.
We show that it should be possible, with the data at hand, to fit the tail of
the muon impact parameter distribution by adopting the chain decay model
proposed in \cite{Giro-small}, $\phi\rightarrow2\phi_{1}\rightarrow4\phi_{2}
$, where $\phi_{1,2}$ are two other SM singlets and $\phi_{2}$ decays into a
tau pair with a lifetime around 30 picoseconds. Our main purpose is to make
possible a close comparison of the model with the data so as to allow a
decisive test. To this end we discuss in Section 4 a few quantitative
predictions of the model that should be compared with the data.

The $\phi$-pair production cross section needed to explain \textit{the signal}
is about 200 pb, in turn requiring a scale of the effective operator that
describes it of about 100 GeV. To make it acceptable, we find it necessary to
\textit{deconstruct} the effective operator in terms of renormalizable
interactions, with new particles mediating the communication between the
quarks in the proton and the \textit{hidden} sector that contains $\phi$.
These new particles and interactions must not lead to any unseen structure in
the multi-muon data and must be consistent with known experimental
constraints. Given the size of the cross section and the low effective scale
involved, this proves nontrivial to achieve. Nevertheless we illustrate in
Section 5 two minimal models that can pass the test, to the best that we can
tell. One involves a scalar exchange in the $s$-channel and another a fermion
exchanged in the $t$-channel. In both cases the couplings to the singlet
$\phi$ saturate perturbation theory. We also briefly comment on the problems
that one may face in turning them into complete satisfactory extensions of the SM.

If this interpretation of the CDF events will resist a further scrutiny along
the lines we are proposing, the existence of the light scalar sector suggests
a connection with recent astro-particle data. In Section 6 we consider a
hidden-sector Dark Matter model where DM annihilations into $\phi$'s that decay into $\tau$'s
can provide the positron excess claimed in $e^{+}$ cosmic rays by
PAMELA~\cite{PAMELA}, while giving in the $e^{+}+e^{-}$ spectrum a feature
somewhat smoother than the peak claimed by ATIC~\cite{ATIC-2}. Conclusions are
summarized in Section 7.

\section{The production of the \textit{signal} events}

\label{Production}

As mentioned, we first concentrate our attention on the events with two cones
containing at least two muons each, which we call \textit{signal} events
as their features are peculiar enough that they should be minimally polluted
by known backgrounds and could therefore be
a quasi-pure sample of a beyond the SM signal.
For an integrated luminosity of 2.1 $\mathrm{fb}^{-1}$ there are about four
thousand such events~\cite{Giro-big}. If they have to arise from $\phi\rightarrow4 \, \tau
\bar{\tau} $, this requires a significant $p\bar{p}\rightarrow\phi\phi$ cross
section, above 100 pb. Furthermore the invariant mass distribution of all
muons contained in these events, Fig. 35a of \cite{Giro-big}, which we aim to
explain, does not show any special feature other than a threshold rise and an
extended tail. In view of this we are led to consider an effective operator
bilinear in $\phi$ and with two gluons or a quark-antiquark pair. There are
three such operators of dimension less or equal to 6\footnote{For
definiteness, we assume that $\phi$ is a complex field.}:
\begin{equation}
\label{eq:operators}O_{5}=\frac{1}{\Lambda}(\bar{q}q)|\phi|^{2},~~O_{6F}%
=\frac{1}{\Lambda^{2}}(\bar{q}\gamma_{\mu}q)\,(\phi^{\ast}\overleftrightarrow
{\partial}_{\mu}\phi),~~O_{6G}=\frac{1}{\Lambda^{2}}G_{\mu\nu}^{a}G_{\mu\nu
}^{a}|\phi|^{2},
\end{equation}
where $q$ is a quark field, either $u$ or $d$, and $G$ the gluon field. We
will normalize our plots below assuming that $O_{5}$ couples only to $u$,
while $O_{6F}$ couples to both $u$ and $d$ with equal strength.

With the amplitude corresponding to each of these operators, we have simulated
the $\phi$-pair production followed by the decay chain
\begin{equation}
\phi\rightarrow2~\phi_{1}\rightarrow4~\phi_{2}\rightarrow8~\tau
\label{eq-decay}%
\end{equation}
with $m_{\phi}=15~$GeV; see~ \cite{Giro-small} and Section \ref{Other} for a
discussion of this choice. Imposing the experimental cuts and detection
efficiencies\footnote{Trigger muons must have transverse momentum $p_{T}>
3\,$GeV, rapidity $|\eta|<0.7,$ and are detected with efficiency
$p_{\text{eff}}=0.44$; additional muons must have $p_{T}>2$GeV, $|\eta|<1.1$
and $p_{\text{eff}}=0.838$. Furthermore, the invariant mass of the trigger
muon pair must be between 5 and 80 GeV; the relative azimuthal angle for the
opposite sign trigger muon pair must be below $3.135$ radians.} we get for
each of the above operators the signal cross section\footnote{Our simulations
were performed by adding the necessary new particles and processes in
\textsc{Pythia} 8.108 \cite{Pythia} and were cross-checked with the help of
home-grown Monte-Carlo programs written in \textsc{Mathematica} (both
stand-alone and passing Les Houches events to \textsc{Pythia} for parton
shower).}:
\[
\sigma_{5} =0.38\,\text{pb}\left(  \frac{200\text{ GeV}}{\Lambda}\right)
^{2},\qquad\sigma_{6F} =1.3\,\text{pb}\left(  \frac{200\text{ GeV}}{\Lambda
}\right)  ^{4},\qquad\sigma_{6G} =0.52\,\text{pb}\left(  \frac{200\text{ GeV}%
}{\Lambda}\right)  ^{4}.
\]
For reference the total $p\bar{p}\rightarrow\phi\phi^{\ast}$ cross sections
(without imposing cuts and efficiencies) are
\[
\sigma_{5,\text{tot}}=220\,\text{pb},\qquad\sigma_{6F,\text{tot}%
}=105\,\text{pb},\qquad\sigma_{6G,\text{tot}}=370\,\text{pb}\qquad\hbox{for
$\Lambda=200$ GeV.}
\]

\medskip

\begin{figure}[ptb]
\begin{center}
\includegraphics[ height=3in] {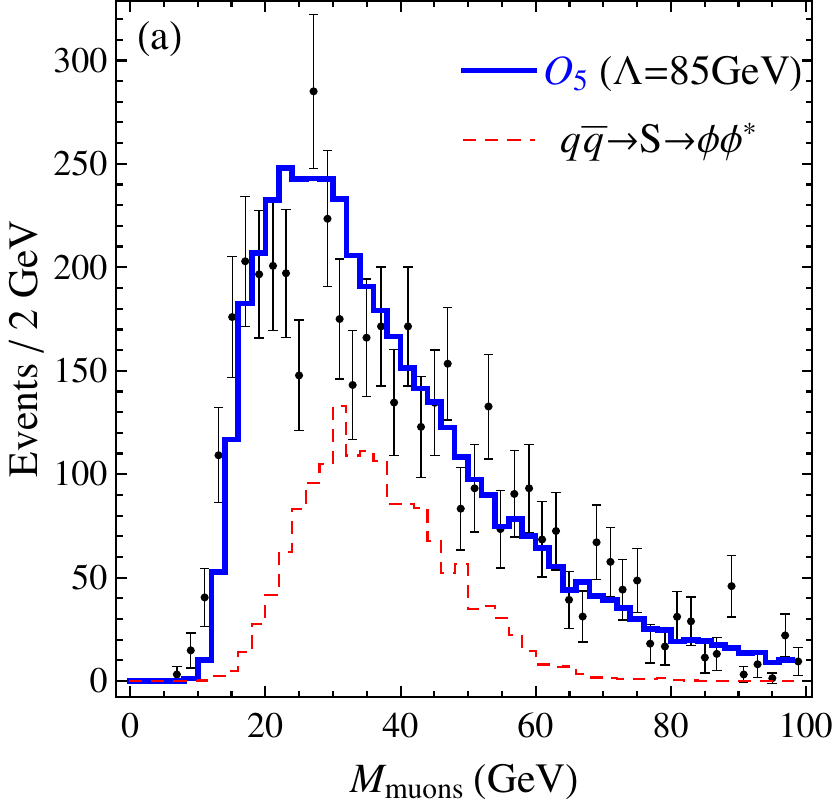} \includegraphics[ height=3in
]{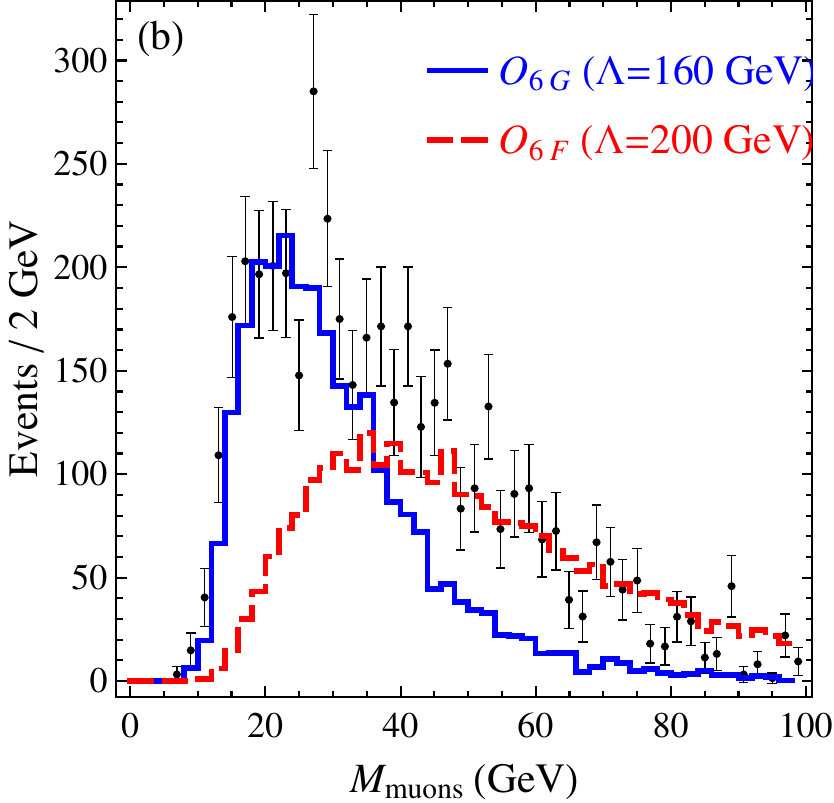}
\end{center}
\caption{\emph{The invariant mass of all muons in events with at least two
muons in both cones. The experimental data are from Fig. 35a of
\cite{Giro-big}, $L_{\text{int}}=2.1$ fb$^{-1}.$}}%
\label{fig-O}%
\end{figure}

Fig. \ref{fig-O} shows how well each of these three production models can
account for the experimental distribution of the invariant mass of all muons
contained in the two signal cones, shown in Fig.~35a of \cite{Giro-big}. The
hardness of the observed distribution motivated us to consider production via
effective non-renormalizable operators. The single relevant parameter here is
the effective scale attached to each individual operator. We find it
remarkable that a single operator, $O_{5}$, can fit the full distribution. For
definiteness we take $q=u$ only, which requires $\Lambda=85~$GeV (fixed in
this case to reproduce the total number of events). On the other hand $O_{6F}$
and $O_{6G}$ can separately reproduce only the tail and the low-mass region
close to threshold, respectively.

The dashed red line in Fig.~\ref{fig-O}a shows the spectrum shape which
results if $\phi\phi^{\ast}$ are decay products of a narrow 300 GeV scalar
resonance: $q\bar{q}\rightarrow S\rightarrow\phi\phi^{\ast}$ (in arbitrary
normalization). The purpose of including this spectrum, clearly unable to
reproduce the experimental data, is twofold. First, this case was considered
in \cite{Giro-small}, and our simulations agree. Second, since the $300$ GeV
$S$-exchange gives a spectrum peaked around $30$ GeV, we can deduce that muons
carry on average around $1/10$ of the total $\phi\phi^{\ast}$ invariant mass.
In particular, events at the tail of the experimental spectrum,
$M_{\text{muons}}\sim100$ GeV, should correspond to $\sqrt{\hat{s}}\sim1$
TeV$.$

\section{Some measured properties of the multi-muons}\label{Other}
Concentrating on the $\phi$-production described by $O_{5}$ with
$\Lambda$ fixed at 85 GeV, we here try to reproduce a few other
significant properties of the muons in the CDF events
of~\cite{Giro-big}: 1) the already mentioned sign-coded multiplicity
distribution; 2) the invariant mass of the muons in a single cone;
3) the muon impact parameter distribution. Notice that data
of~\cite{Giro-big} about 1) and 3) also contain events outside the
sub-sample discussed in Section \ref{Production} that we view as
signal events; thereby we allow for the presence of a background.

\subsection{Sign-coded muon multiplicity}

\begin{figure}[ptb]
\begin{center}
\includegraphics[height=3in] {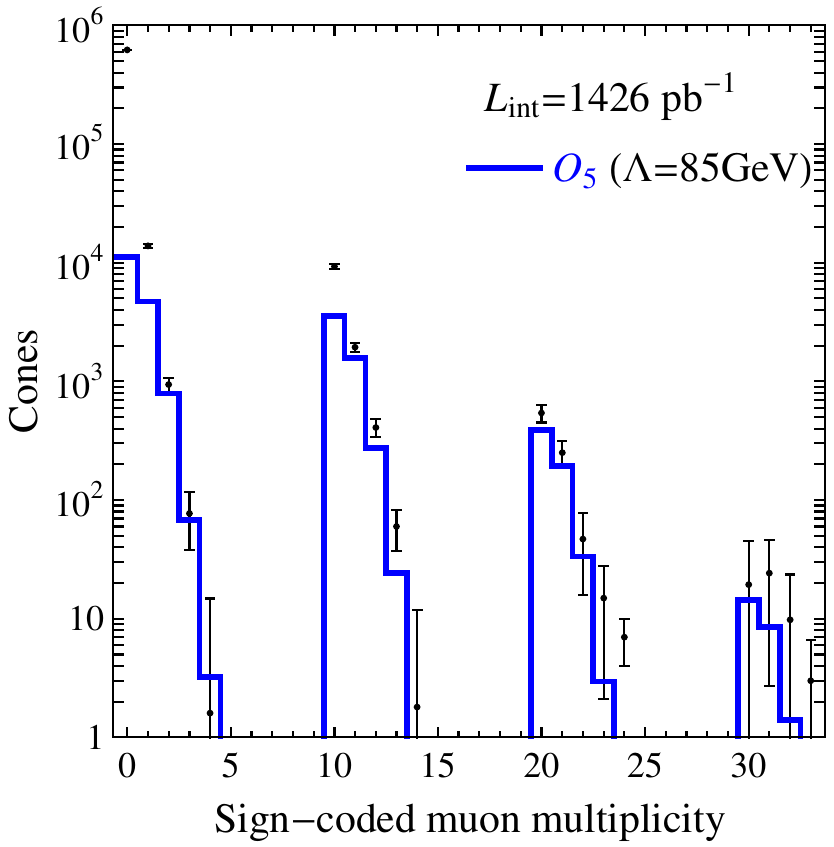}
\end{center}
\caption{\emph{Comparison of our production model with the experimental data
(Fig.~22b of \cite{Giro-big}, significant part only). Every cone enters
separately into this histogram (thus every event enters twice).}}%
\label{fig-mult}%
\end{figure}

For reasons of graphical illustration, the sign-coded muon multiplicity
$\mathcal{M}$ is defined for each of the two $36.8^{\circ}$ cones around the
trigger muons by the formula%
\[
\mathcal{M}=N_{OS}+10N_{SS}%
\]
where $N_{OS}$ $(N_{SS})$ is the number of additional muons in the cone having
opposite (same) sign as the trigger muon. Thus e.g. $\mathcal{M}=0$
corresponds to cones without additional muons. In Table 1 and
Fig.~\ref{fig-mult} we compare the data with our simulation.

\begin{table}[ptb]%
\[%
\begin{array}
[c]{lccc}%
\mathcal{M} & \text{Cone content} & \text{Experiment} & O_{5}(\Lambda
=85\text{GeV})\\
00 & \pm & 620307\pm3413~~ & 11184\\
01 & \pm\mp & 13880\pm573~~ & ~4722\\
02 & \pm\mp\mp & 941\pm135 & ~~790\\
03 & \pm\mp\mp\mp & 77\pm39 & ~~~68\\
04 & \pm\mp\mp\mp\mp & ~1.6\pm13.2 & ~~~~3\\
10 & \pm\pm & 9312\pm425~ & ~3580\\
11 & \pm\pm\mp & 1938\pm173~ & ~1573\\
12 & \pm\pm\mp\mp & 409\pm71~ & ~~277\\
13 & \pm\pm\mp\mp\mp & 60\pm23 & ~~~24\\
20 & \pm\pm\pm & 542\pm91~ & ~~392\\
21 & \pm\pm\pm\mp & 251\pm61~ & ~~194\\
22 & \pm\pm\pm\mp\mp & 47\pm31 & ~~~33\\
23 & \pm\pm\pm\mp\mp\mp & 14.9\pm12.8 & ~~~~3\\
30 & \pm\pm\pm\pm & 19.4\pm25.6 & ~~~14\\
31 & \pm\pm\pm\pm\mp & 24.2\pm21.5 & ~~~~8\\
32 & \pm\pm\pm\pm\mp\mp & ~9.8\pm13.8 & ~~~~1
\end{array}
\]
\caption{\emph{Numerical content of Fig.~\ref{fig-mult}. Experimental data
taken from Table X of \cite{Giro-big} (significant part only). The data
correspond to $L_{\text{int}}=1426$ pb$^{-1}$. The statistical errors of our
simulation ($\sim0.2\sqrt{N_{\text{bin}}}$) are negligible compared to the
experimental errors.}}%
\end{table}

\smallskip

The shape of this distribution is largely independent of the
production mechanism, and indeed our Fig.~\ref{fig-mult} is quite
similar to Fig.~1 of \cite{Giro-small}. Since we have fixed the
production mechanism (the $O_5$ operator), we can simulate not only
the shape but also the absolute normalization ($\Lambda = 85\,{\rm
GeV}$), while in Fig.~1 of \cite{Giro-small} normalization has been
fixed arbitrarily. We see in particular that we can reproduce
essentially all events (within errors) with 3 or more muons in a
cone: only a relatively small systematic excess of not understood
backgrounds is here possibly needed to fully account for the event
rates. However, only about a third of events with 2 muons in a cone
is reproduced. Since we fixed the normalization by using the events
with two or more muons in \textit{both} cones, this shows that
events with only 2 muons in one cone and 1 muon in the other are
still significantly contaminated by backgrounds.

The fraction of events without additional muons that we are able to reproduce
is negligible. Almost all of these events must be background. Indeed according
to \cite{Giro-big} at least 50\% of events without additional muons can be due
to in-flight decays of kaons and hyperons producing real muons or pion punchthroughs.

\begin{figure}[ptb]
\begin{center}
\includegraphics[height=2in] {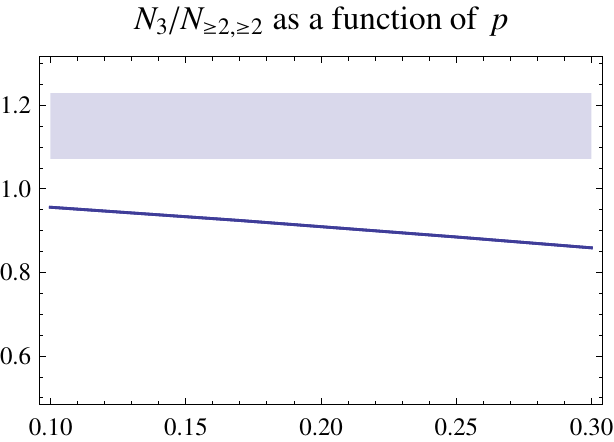}
\includegraphics[height=2in] {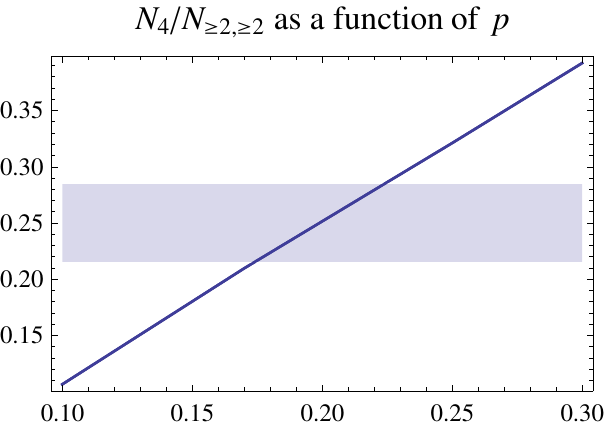}
\end{center}
\caption{\emph{The numbers of events with 3 muons in a cone $N_3$
(the sum of bins 02,11,20 in Table 1) and with 4 muons in a cone
$N_4$ (the sum of bins 03,12,21,30) relative to the number of events
with at least two muons in each cone $N_{\ge2,\ge2}$, simulated as a
function of the varying $\tau\to\mu$ branching ratio $p$. The bands
show experimental values. The physical value $p=0.17$ gives
reasonable agreement in both plots. One can see that the ratio
$N_3/N_{\ge2,\ge2}$ is rather insensitive to the variation of $p$,
while the ratio $N_4/N_{\ge2,\ge2}$ prefers the values of
$p\sim0.2$, close to the physical value.}}
\label{fig-P}%
\end{figure}

Notice that the above agreement of simulation with experiment, in
bins with $\geq3$ muons, relies on the precise value of $p={\rm
BR}(\tau\rightarrow\mu )\approx0.17$, thereby indirectly supporting
the $\tau$ interpretation. This can be seen by generating toy
Monte-Carlos in which $p$ varies in the $0.1\div0.3$ range (while
keeping fixed the muon detection efficiencies). The number of events
with 3 muons in a cone (including sign-coding), relative to the
events with at least two muons in both cones, is largely independent
of $p$. This is not unexpected, since the probability to put 2
additional muons in the same cone is roughly the same as to put them
in different cones. On the contrary, the relative number of events
with $4$ or more muons in a cone was quite sensitive to $p$, to the
extent that $p=0.1(0.3)$ led to clear shortage(excess) of events in
these bins, compared to experiment, see Fig.~\ref{fig-P}. At the
same time the relative distribution of these events into the various
sign-coded bins is determined by simple combinatorics and, as a
result, proves to be insensitive to the variations of $p$.

\subsection{Single cone muon invariant mass}

\begin{figure}[ptb]
\begin{center}
\includegraphics[width=3in] {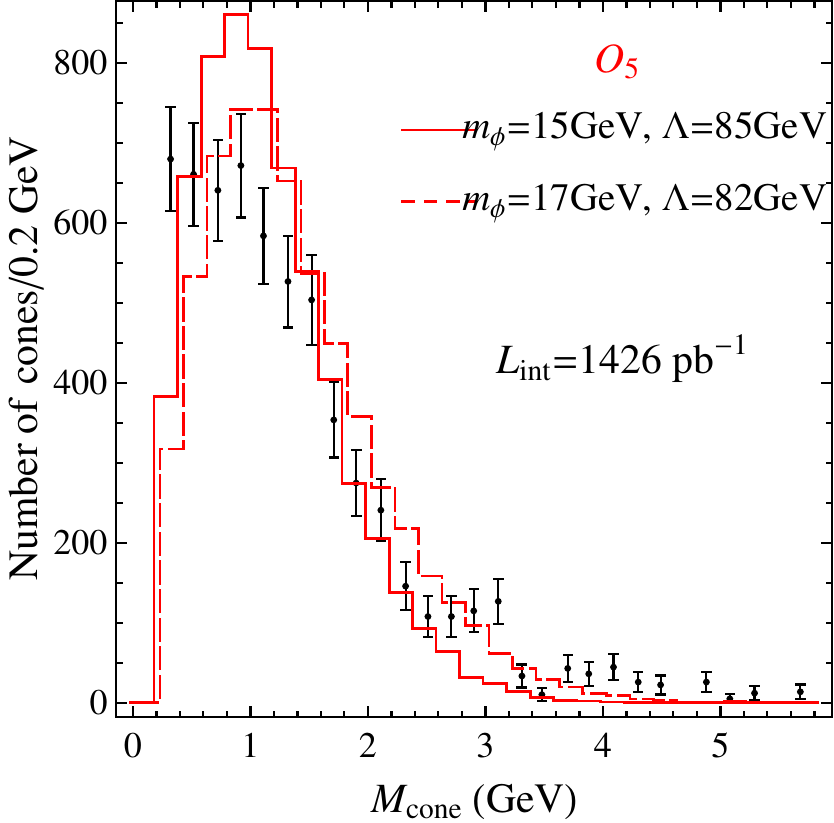}
\end{center}
\caption{\emph{Invariant mass\thinspace, $M_{\text{cone}}$, of all muons in a
cone when both cones contain at least two muons. Data reproduced from Fig.~41a
of \cite{Giro-big}.}}%
\label{fig-cone}%
\end{figure}

The multi-muon events are characterized by low invariant mass of muons in a
single cone, with a spectrum which sharply cuts off at $2\div3$ GeV. It was
shown in \cite{Giro-small} that this cutoff can be reproduced within the
hypothesis of the decay chain (\ref{eq-decay}) if the mass of $\phi$ is not
much above the $8\tau$ threshold ($14.2$ GeV). This conclusion is largely
independent of the production mechanism, and in particular holds within our
model. In Fig.~\ref{fig-cone} we compare the data with our simulation for
$m_{\phi}=15$ and $17$ GeV (for events where both cones contain at least 2
muons, i.e. the \textit{signal} events of Section \ref{Production}).

The overall normalization is fixed in both cases to reproduce the total number
of events. While the data are reproduced in an acceptable way for these two masses,
higher values (see \cite{Giro-small} for $m_{\phi}=20$ GeV) start showing a
clear deviation.

We however see that a discrepancy is present at small
$M_{\text{cone}}$: while the data remain almost constant down to the
lowest bin, the simulation drops there. The physical reason for this drop is
intuitively clear: it corresponds to a small probability to have an almost
collinear muon pair.
Without a precise understanding of the detector (e.g.\ possible punch-through of hadrons mimicking muons)
and of the statistical correlations
between the various bins we cannot say more on this issue nor perform a precise $\chi^2$ analysis.

%This feature of the data can perhaps be due to some
%detector effect.

\medskip

It is legitimate to ask if the experimental distribution in
Fig.~\ref{fig-cone} also contains a peak corresponding to the
possible subdominant $\phi_{2}$ decay channel, directly into muons,
$\phi _{2}\rightarrow\mu^{+}\mu^{-}$. Such a peak would have to be
located at $m_{\phi_{2}}$, i.e.\ slightly above
$2m_{\tau}\simeq3.55$ GeV\footnote{And not at $7.2$ GeV as in Fig.~1
of \cite{Matt}.}$.$ Indeed we see some excess of events around this
value of $M_{\text{cone}}$ (which could become statistically
significant with more statistics), but, at the same time, we do not
forget an important source of dimuon pairs --- $c\bar{c}$ mesons ---
present in the $3\div4$ GeV mass region. From the model-building
point of view, the branching ratio
$\phi_{2}\rightarrow\mu^{+}\mu^{-}$ may well be suppressed to an
unobservable level (see Section \ref{Dec}).

\subsection{The muon impact parameter}

\begin{figure}[ptb]
\begin{center}
\includegraphics[width=2.5in] {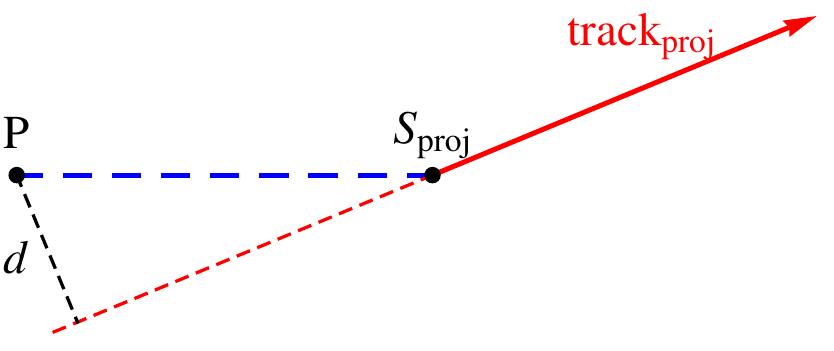}
\end{center}
\caption{\emph{The impact parameter $d$ is the distance between the track
projection on the transverse plane $\mathrm{track}_{\mathrm{proj}}$ and the
primary interaction vertex P, which is experimentally known with high
precision (in particular due to many soft hadronic tracks originating from
it), while the secondary vertex S is not measured unless more than one track
originates from it.}}%
\label{fig-d-expl}%
\end{figure}

For a given charged particle track, the impact parameter $d$ is defined as the
distance between the track and the primary interaction vertex in the
transverse plane, see Fig.~\ref{fig-d-expl}.

One of the most striking properties of multi-muon events is that the impact
parameter distribution of muon tracks has an exponential tail. Such a tail can
be explained if the tracks originate from decays of a long-lived particle.
Notice that a boost $\gamma$ of the decaying particle does not affect the
typical value of the impact parameter, since it increases the typical decay
distance $\ell\sim\gamma c\tau_{0}$ but also decreases the typical collimation
angle of decay products by the same factor $\gamma$. As a result the decay
scale of the impact parameter distribution is of the order of the decaying
particle's $c\tau_{0}$. The precise proportionality coefficient in general
depends on the kinematic cuts.

\begin{figure}
\begin{center}
\includegraphics[width=3in] {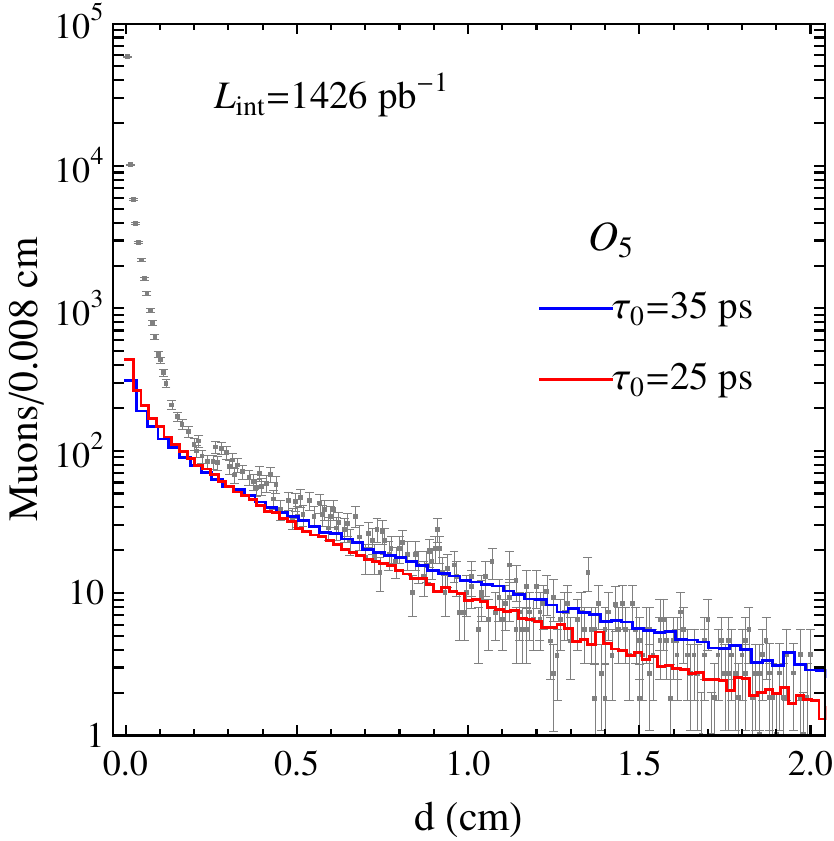}\qquad\includegraphics[width=3in] {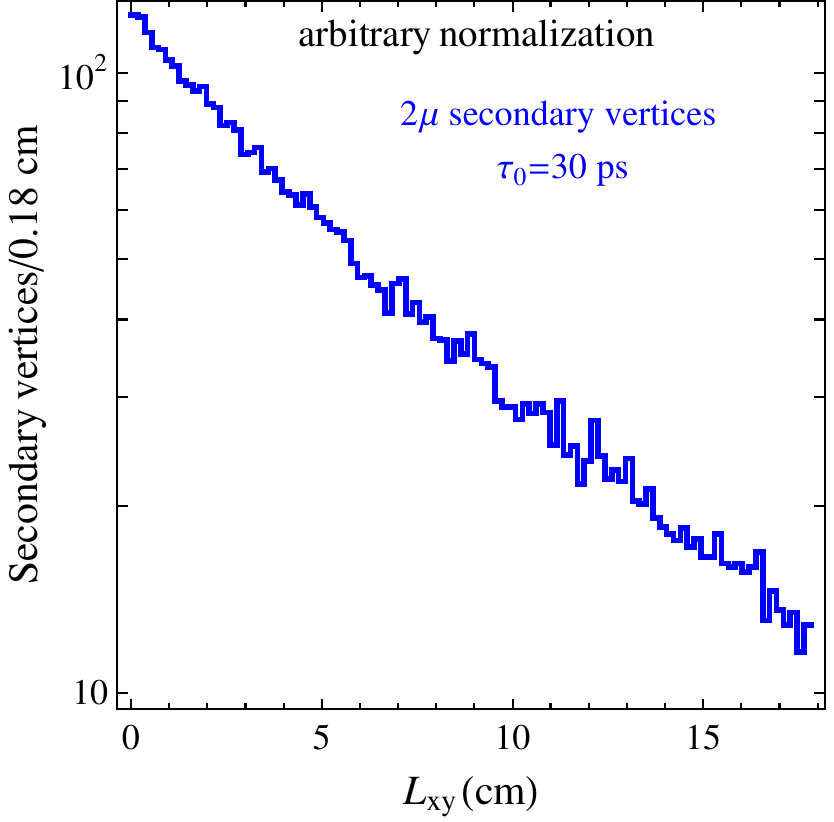}
\end{center}
\parbox{0.5\textwidth}{
\caption{\emph{Primary muon impact parameter distribution for cones containing
exactly two muons. Data taken from Fig.~25a of \cite{Giro-big}.}}%
\label{fig-impact}}
\parbox{0.5\textwidth}{
\caption{\textit{The simulated distribution for }$L_{xy}$\textit{ for }%
$\tau_{0}=30$\textit{ ps and for dimuon secondary vertices (in arbitrary
normalization). }}%
\label{fig-lxy}}
\end{figure}

In Fig.~\ref{fig-impact} we compare one of the several published impact
parameter distributions (impact parameters of primary muons of all cones
containing exactly 2 muons) with our simulation. We assume that $\phi
\rightarrow2\,\phi_{1}\rightarrow4\,\phi_{2}$ decays are prompt, while
$\phi_{2}\rightarrow\tau\bar{\tau}$ has a long lifetime. As discussed in
\cite{Giro-small}, this assumption is forced on us by another piece of
experimental data which we do not discuss here: no significant correlation
between the impact parameters of different muonic tracks in the same cone.

We see that both the decay rate and the normalization of the exponential tail
can be reproduced for a $\phi_{2}$ lifetime around 30 ps.\footnote{This value
has to be taken with a grain of salt since the tail of the experimental
distribution may be significantly contaminated by $K_{S}^{0}$ in-flight decays
(lifetime 90 ps).} Note that `naive' decay exponent would give a 20 ps
lifetime; the fact that the naive ultra-relativistic estimate underestimates the true lifetime by
30\% in this particular case has already been noticed in \cite{Giro-small} on
the basis of a toy simulations. We also see that the large number of events at
small $d$ are not reproduced by our signal model.
Most of these events must be
due to QCD contributions, e.g.\ sequential $b$ quark decays which had been
subtracted in the distributions considered previously but not in Fig.~\ref{fig-impact}.
All data are extracted from~\cite{Giro-big}, so that we follow their definitions.

%, which are not
%subtracted in this plot (unlike in the sign-coded multiplicity distribution of
%Fig.~\ref{fig-mult}).

\section{Possibilities for further experimental tests}

Now that a concrete proposal for the $\phi$-production via the $O_{5}$
operator has emerged, what other possible tests could provide further checks
of the model? Many such tests can be done using information about hadrons
present in the multi-muon events. However, any study with hadrons would
require detector simulation, especially because the relevant experimental
information \cite{Giro-big} involves hadronic \textit{tracks} rather than,
say, isolated jets.

In this Section we discuss 3 further experimental tests which involve only
muons and which might be relatively easier to implement. The corresponding
experimental distributions are not yet available. These tests are: 1) dimuon
displaced vertices; 2) muon track alignment; 3) deviation from
back-to-backness in the CM\ frame of the hard process. \textit{For
illustrative purposes only} we will show in all cases the expected
distributions from our simulation.

 All our predictions below are computed using our simplified
simulation of the triggering and reconstruction process, see
footnote 1. In order to perform a precise comparison with the data,
the predictions should be recomputed using the full simulation of
the detector, available only to the experimental collaborations.

\subsection{Dimuon displaced vertices}

If the model is right, a fraction of events must contain a displaced
secondary vertex with two muon tracks originating from it. Such a
vertex occurs when both taus in $\phi_{2}\rightarrow\tau\bar{\tau}$
decay into a muon. According to our simulation, for
$L_{\text{int}}=2.1$ fb$^{-1}$ there will be about 2000 such
secondary vertices, for $\phi_{2}$ lifetime of 30 ps almost all of
them within the CDF inner silicon vertex detector ($10.6$ cm
radius). Taking into account excellent tracking capabilities of the
CDF detector, a significant fraction of these vertices should be
identifiable.

A useful experimental quantity to describe the secondary vertex
distribution is called $L_{xy}$, the distance between the secondary
and primary event vertices projected onto the transverse momentum of
the two-track system. In Fig.~\ref{fig-lxy} we give our simulated
distribution for $L_{xy}$ when the secondary vertex is determined by
two muon tracks.

\subsection{Muon track alignment}

As discussed in Section \ref{Other}, $m_{\phi}$ cannot be much above
$15\div17$ GeV. On the other hand we must have $m_{\phi_{2}}>2m_{\tau}$
otherwise the decay of $\phi_{2}$ will proceed via off-shell taus with a
typical lifetime much bigger than the needed 30 ps. One consequence of such
restricted kinematics is that the four $\phi_{2}$ are non-relativistic in the
rest frame of the parent $\phi,$ with typical kinetic energy%
\[
E_{\text{kin}}=\frac{m_{\phi}}{4}-m_{\phi_{2}}\simeq(0.2\div0.7)\,\mathrm{GeV}%
.
\]
As noticed in \cite{Matt}, this leads to $\phi_{2}$ decay vertices collinearly
aligned along the $\phi$ momentum. Indeed, their typical separation $\delta$ in the
orthogonal direction is
\begin{equation}
\delta \sim\beta\cdot c\tau_{0}\text{,\qquad}\beta\sim(2E_{\text{kin}}/m_{\phi_{2}%
})^{1/2}\simeq0.3\div0.6\,.\label{eq-typ}%
\end{equation}

Since we are here concerned only with muon tracks, the $\phi_{2}$ decay
vertices cannot be reconstructed in most cases. However, a nontrivial
experimental measure of their collinear alignment can be defined purely in
terms of muonic tracks, for events with at least 3 muons in a cone. We will
use the following quantity:%
\[
A=\min_{\vec{p}}\left(  \sum_{i}[d_{i}(\vec{p})]^{2}\right)  ^{1/2}\,
\]
where $d_{i}$ is the distance between the track of the $i$-th muon and a
straight line passing through the primary vertex in the direction $\vec{p}$.
By tracks here we mean idealized reconstructed tracks, where the effects of
bending in the magnetic field of the detector have been deconvoluted. Thus
$A=0$ exactly when all tracks intersect a single line pointing to the primary
vertex. Generically we expect $A$ positive and of the order of
Eq.~(\ref{eq-typ}), although this is likely to be an overestimate since the
definition of $A$ makes it automatically vanish for cones with only two muons.

\begin{figure}[t]
\begin{center}
\includegraphics[width=3in] {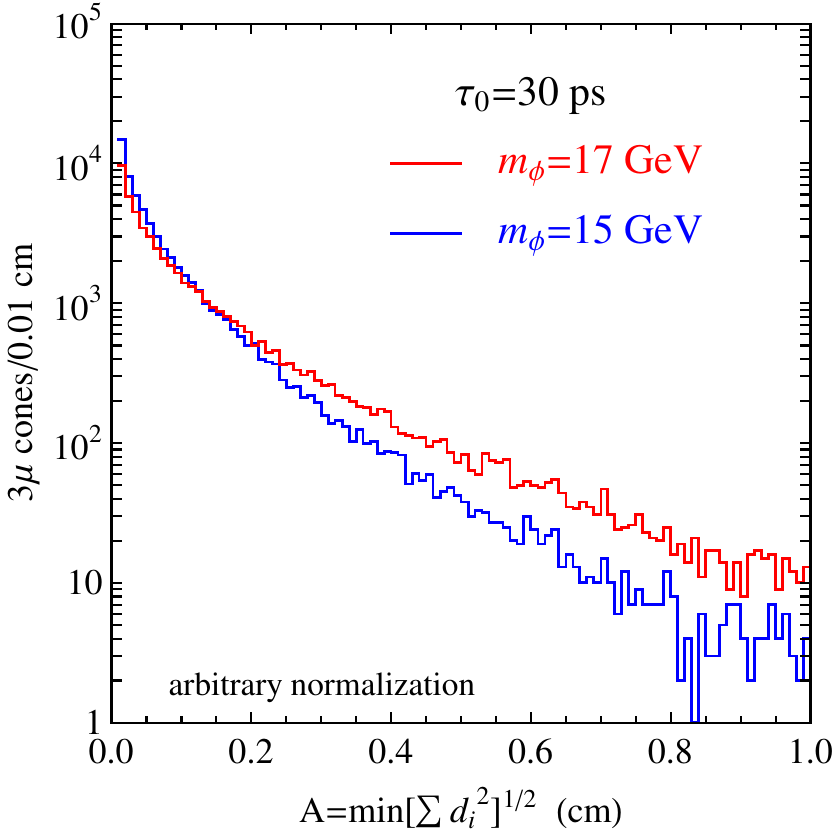}\qquad
\includegraphics[width=3in] {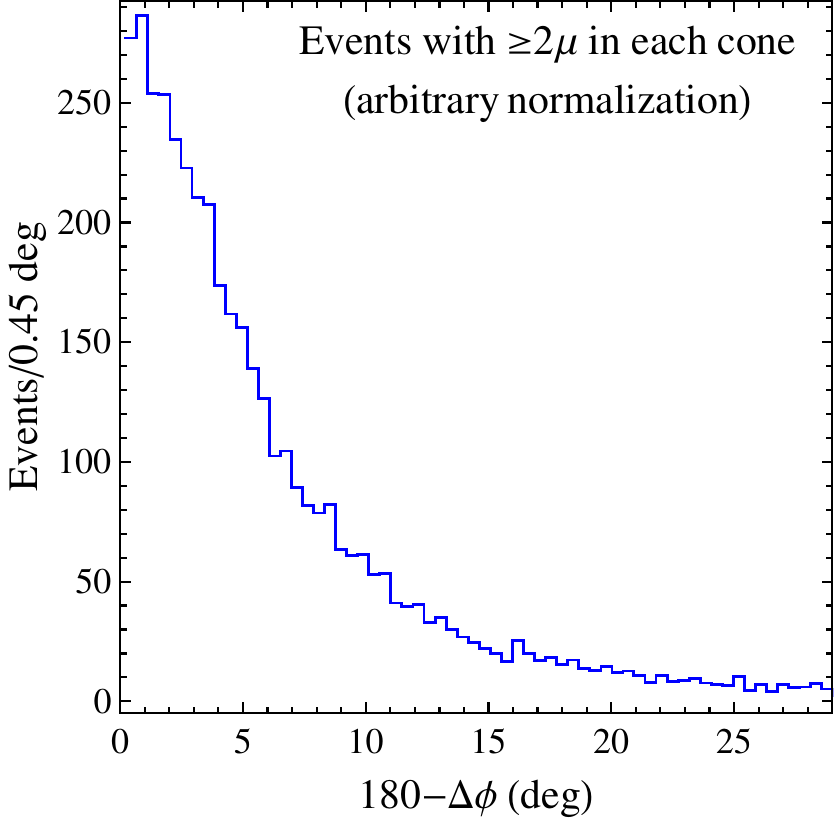}
\end{center}
\parbox{0.5\textwidth}{
\caption{\emph{Simulated distributions of the muon track alignment parameter
$A$, for events with exactly 3 muons in a cone.}}%
\label{fig-align}}
\parbox{0.5\textwidth}{
\caption{\emph{The expected `deviation from back-to-backness' angle
$\pi-\Delta\varphi$ distribution for events with at least 2 muons in each
cone.}}%
\label{fig-bb}}
\end{figure}

In Fig.~\ref{fig-align} we plot the simulated distribution of $A$ of events with at least 3 muons in a cone.
A less steep distribution, similar to the one of the impact parameter $d$ in Fig.~\ref{fig-impact},
would instead arise from non-aligned $\tau$ decays.
In Fig.~\ref{fig-align} we assumed two values of the $\phi$ mass: $15$ and $17$ GeV, $m_{\phi_{2}}=3.6$ GeV
and  $m_{\phi_{1}}=7.3$ GeV.%
%{\bf\color{red} I REMOVED THE DISCUSSION OF THE AVERAGE A BECAUSE a) THIS AVOIDS DISCUSSING
%WHY THE APPROX IS NOT VERY GOOD, DON'T CARE AND LOOK THE FULL FIGURE
%b) GIVEN THE BACKGROUNDS THE BEST OBSERVABLE IS THE SLOPE RATHER THAN THE AVERGE A,
%AND THE REFEREE CAN COMPARE WITH THE SLOPE OF $d$}

%The corresponding average values are:%
%\[
%\left\langle A\right\rangle =0.07,\,0.11\,\mathrm{cm},
%\]
%respectively.
%The tendency of increase in $\left\langle A\right\rangle $ is
%not fully consistent with $\sim\sqrt{E_{\text{kin}}}$ as (\ref{eq-typ}) would
%suggest. We assign this discrepancy to the possible role of $m_{\phi_{1}}$,
%which was fixed at $m_{\phi_{1}}=7.3$ GeV in the above simulation.

\subsection{Deviation from back-to-backness}

While the total muon momenta in each cone $\vec{p}_{1,2}$ is only a fraction
of their parent $\phi$ momentum, so that the muon invariant mass gives a poor
information on the $\phi$ mass, $\vec{p}_{1}$ is very strongly correlated to
the direction of $\phi$, allowing to test details of the $\phi$ production mechanism.

The distribution of $\phi$'s in rapidity does not contain much information,
being mostly shaped by the experimental cuts and by the parton distributions.
More interesting is the distribution of the relative azimuthal angle
$\Delta\varphi$ (i.e.\ the relative angle between $\vec{p}_{1}$ and $\vec
{p}_{2}$ in the plane transverse to the beam), since the deviation from the
back-to-back configuration ($\Delta\varphi=\pi$) can be due only to QCD
radiation in our model. In our model $\phi$ is a color singlet, so that only
Initial State Radiation is present, leading to small deviations from
back-to-backness. Fig.~\ref{fig-bb} shows our result for the simulated
distribution of $\Delta\phi$ for signal events containing at least two muons
in each cone: the average angle is
\[
\left\langle 180^{\circ}-\Delta\varphi\right\rangle \approx7^{\circ} .
\]
We relied on \textsc{Pythia} 8.108 with its standard settings of ISR
parameters and no detector simulation, and we have not attempted to quantify
QCD uncertainties on this result. Once these issues are settled, such
distribution can discriminate from other production mechanisms or possible
backgrounds. For instance, $\phi$ production via the alternative gluon
operator $O_{6G}$ gives $\sim2$ times wider distribution since Initial State
Radiation off gluons is stronger. A wider distribution would also be obtained
in presence of \textit{Final} State Radiation, which could be the case if the
multi-muon events are background associated with dijet events. Indeed,
analogous experimentally measured distributions for $b\bar{b}$ production
\cite{bbar} are significantly wider.

\section{Explicit realizations for $O_{5}$}

\label{Dec}

We are not concerned here by the nature of the \textit{hidden} world, and we
do not pretend to \textit{explain} the interaction that leads to the last step
of the chain decay: $\phi_{2}\rightarrow\bar{\tau}\tau$. Given its small
width, we can attribute it to another effective operator
\begin{equation}
L_{\text{decay}}=\frac{1}{\Lambda^{\prime}}\bar{l}_{3}\tau\,H_{s}^{c}\phi_{2}%
\end{equation}
where $H_{s}$ is the standard Higgs field, ${l}_{3}$ is the third-generation
lepton doublet, $\tau$ the singlet. The scale $\Lambda^{\prime}$ associated
with it is much above the Fermi scale, unlike the $\Lambda$ associated to
$\phi$-production.

On the contrary, to be credible, the effective non-renormalizable operators in
eq.~(\ref{eq:operators}) suppressed by a scale $\Lambda$ as low as one hundred
GeV must be \textquotedblleft deconstructed" in terms of some explicit
particle exchanges. What can mediate the pretty strong communication between
the SM particles and the hidden world represented by $\phi$ or the other light
particles to which $\phi$ decays, without running in conflict with known
experimental constraints?

We concentrate our attention to $O_{5}$, leaving the two dimension-6 operators
to Appendix A.

There are only two ways in which the operator $O_{5}$ can be generated by the
mediation of a single particle: 1) a scalar exchanged in the $s$-channel, with
the quantum numbers of the standard Higgs doublet, but different from it and
with a negligibly small vacuum expectation value; 2) a colored fermion
exchanged in the $t$-channel. We analyze both cases in turn.

\subsection{Scalar exchange}

The renormalizable Lagrangian that gives rise to $O_{5}$ after integrating out
the heavy scalar doublet $H$, with hypercharge $-1/2$, is
\begin{equation}
{L}_{H}=\lambda_{q}\,\bar{q}uH+\lambda\,H^{\dagger}H_{s}|\phi|^{2}-M_{H}%
^{2}H^{\dagger}H\,, \label{LH}%
\end{equation}
where $q$ is the first generation left-handed quark doublet and $u$ is the
right-handed up quark. (Similar considerations would hold for the down quark.)
In this way, at energies below $M_{H}$, one generates an effective Lagrangian
with $O_{5}$ for the up quark and a scale
\begin{equation}
\Lambda=\frac{M_{H}^{2}}{\lambda_{q}\lambda v},
\end{equation}
which, to match with Fig.~\ref{fig-O}, should be about 100 GeV.

Requiring as the perturbativity limit that the partial width $H\rightarrow
\phi\phi^{\ast}$ be less than the $H$ mass\footnote{For $M_{H}<3$ TeV this
turns out to be stronger than the NDA bound $\lambda<16\pi^{2}$.}, we get%
\begin{equation}
\Gamma(H\rightarrow\phi\phi^{\ast})\sim\frac{1}{16\pi}\frac{(\lambda v)^{2}%
}{M_{H}}<M_{H}\qquad\Longrightarrow\qquad\lambda v<\sqrt{16\pi}M_{H} .
\label{l1}%
\end{equation}
From the last two equations we conclude that%
\begin{equation}
M_{H}/\lambda_{q}\lesssim700\text{ GeV.} \label{limit}%
\end{equation}
This bound applied to the Lagrangian (\ref{LH}) raises issues of potential
conflicts with experiments, which we briefly address in Section 5.3. Here we
focus on its possible effects at the Tevatron itself.

\begin{figure}[ptb]
\begin{center}
\includegraphics[width=3in] {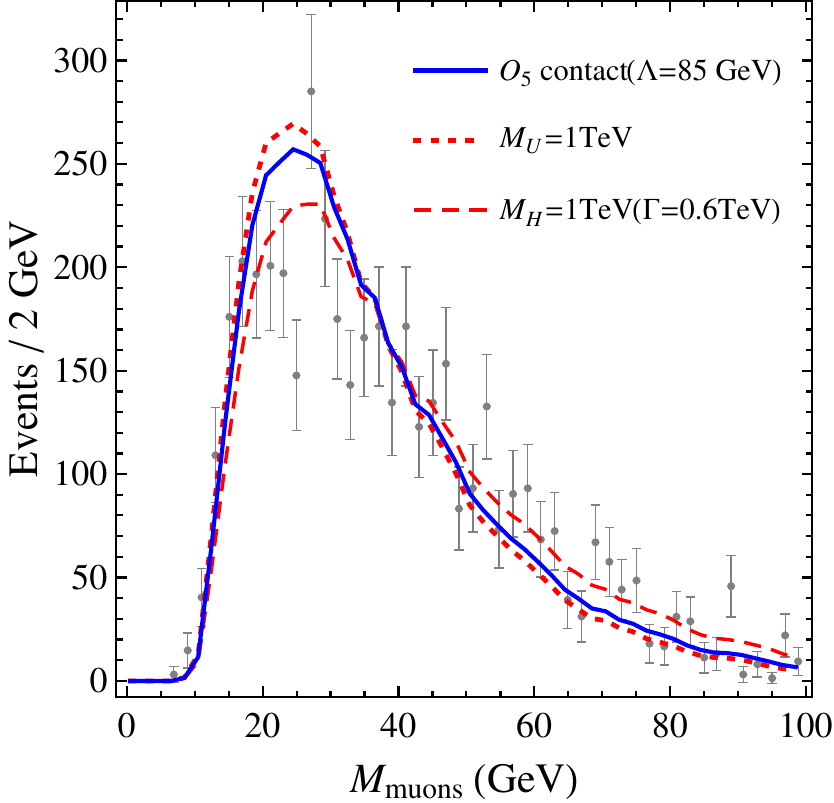}
\end{center}
\caption{\emph{The analogue of Fig.~\ref{fig-O}a where the contact term
interaction generated by the local operator $O_{5}$ is compared with a scalar
$s$-channel effect and a fermion $t$-channel exchange generating the same
operator at low energies. The couplings are slightly adjusted so that the
total number of events is the same in all cases.}}%
\label{fig-q2}%
\end{figure}

The first of these effects concerns the same $p\bar{p}\rightarrow\phi
\phi^{\ast}$ production, since the effective-operator approximation in
describing the $H$-exchange may be invalid. This is illustrated in
Fig.~\ref{fig-q2} where the effective-operator result of Fig.~\ref{fig-O}a is
compared with the exchange of the $H$-scalar with a mass of 1 TeV and a width
of 600 GeV. For these values of the parameters we see no conflict with the
data, which may even be better described in the second case.

\smallskip

Another effect of the $H$-exchange might have shown up in dijet properties.
CDF has published studies both of the angular distribution of the dijets
\cite{CDFcomp} and of a possible resonant structure in the dijet mass spectrum
\cite{CDFdijet}. In both cases it is nontrivial for us to compare these
results with the signal we might expect. For the case of the angular
distribution, the $H$-exchange generates the operator
\begin{equation}
-\frac{\lambda_{q}^{2}}{2M_{H}^{2}}(\bar{u}u)^{2}, \label{o2}%
\end{equation}
whereas the analysis in \cite{CDFcomp} concerns the operator (including the
sign)%
\begin{equation}
-\frac{1}{2\Lambda^{2}}(\bar{q}_{L}\gamma^{\mu}q_{L})^{2},\quad q=(u,d)^{T}%
,\quad\Lambda>700\text{ GeV at 95\% C.L.\thinspace.} \label{o1}%
\end{equation}
The analysis leading to this bound is based on the fact that operator
(\ref{o1}) interferes constructively with the QCD $q\bar{q}\rightarrow
q\bar{q}$ scattering amplitude, leading to an increase of events at large
scattering angles. Our preliminary analysis shows that the interference term
between the operator (\ref{o2}) and the QCD amplitude is smaller. Thus we
expect that the bound in (\ref{limit}) should not pose a problem. All this is
preliminary, however, also because the resonance may be within the accessible
range. Indeed the same resonance search in \cite{CDFdijet} could be of
importance to us. The problem in this case, however, is that the published
limits concern only narrow resonances, when the width is within the
resolution, while we are dealing with the opposite case. How to rescale from
this, if possible at all, is to the least nontrivial. We believe however that
the current results may only constrain $M_{H}$ weakly enough to allow
consistency with our effects in a significant range of the parameters.

\subsection{Fermion exchange}

\begin{figure}[ptb]
\begin{center}
\includegraphics[height=1in] {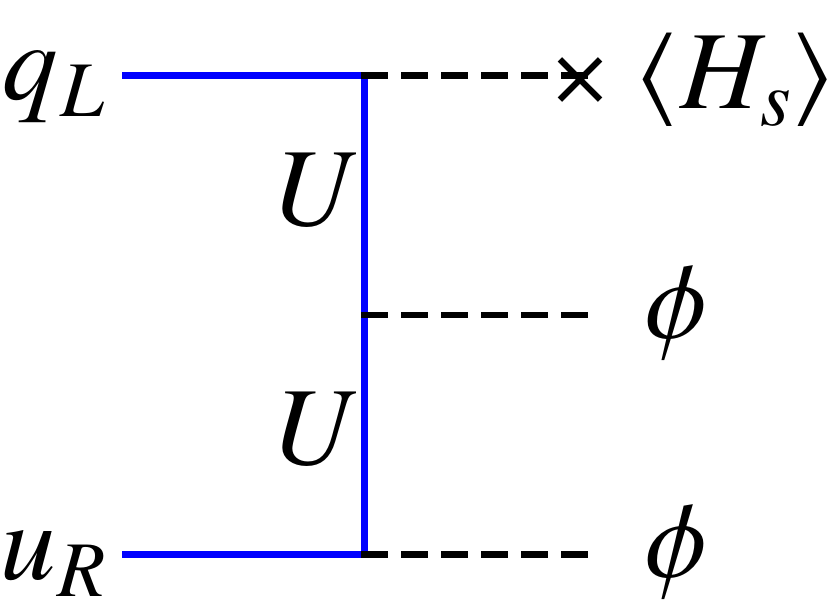}\qquad
\includegraphics[height=1in] {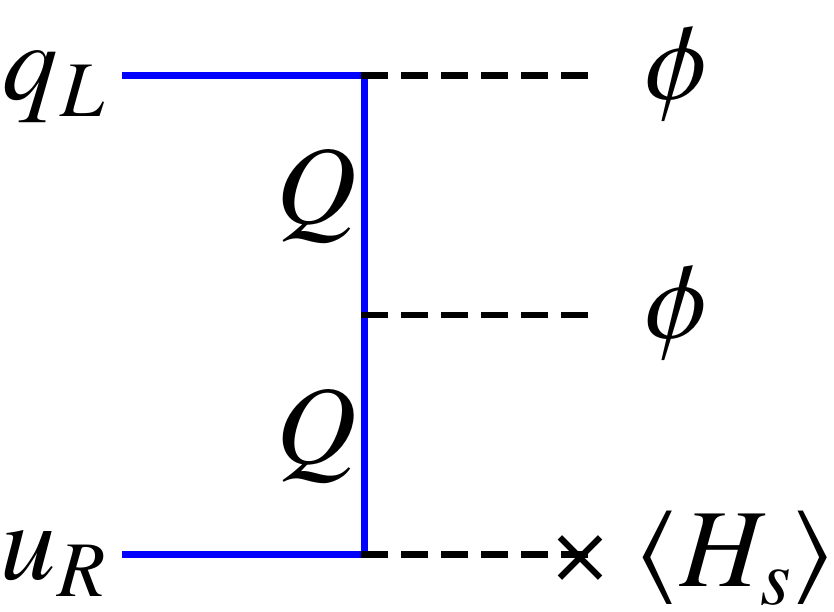}\qquad
\includegraphics[height=1in] {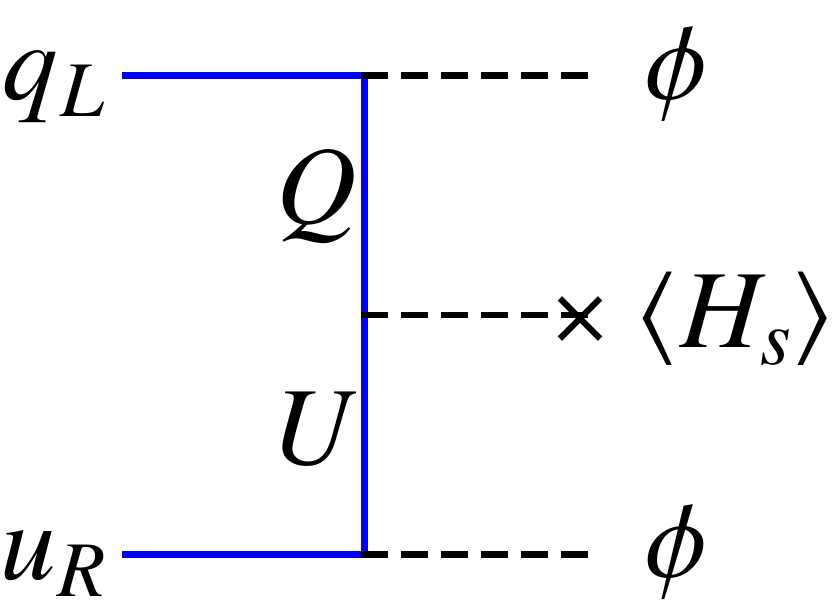}
\end{center}
\caption{\emph{Three ways to generate $O_{5}$ via new heavy quark exchange in
the t-channel. Only the last way, corresponding to the Lagrangian (\ref{LF}),
is phenomenologically acceptable.}}%
\label{fig-ferm}%
\end{figure}

The other way to generate the operator $O_{5}$ is by a heavy quark exchange in
the $t$-channel, see Fig.~\ref{fig-ferm}. In principle one can do it with one
SU(2)$_{L}$-singlet $U$ or with one doublet $Q$, but this requires large
couplings of the form
\begin{equation}
\lambda_{q}\bar{q}UH_{s},~~\lambda_{q}\bar{Q}uH_{s},
\end{equation}
respectively for the $U$ or the $Q$ cases (here $q$ is the standard SM
left-handed doublet). 

However, after diagonalization of the full mass matrix in the 2/3-charge
sector, this would lead to a large wrong component of the physical
$u_{L}^{\text{ph}}$ or $u_{R}^{\text{ph}}$, which is not acceptable. One needs
therefore to introduce both a $U$ and a $Q$ and the Lagrangian
\begin{equation}
{L}_{F}=(\lambda_{q}\bar{q}Q\phi+\lambda_{u}\bar{U}u\phi^{\ast}+\lambda\bar
{Q}_{R}U_{L}H_{s}+h.c.)+M_{Q}\bar{Q}Q+M_{U}\bar{U}U, \label{LF}%
\end{equation}
which, in the local limit and the approximations $M_{U}\approx M_{Q}\gg
\lambda~v$, gives rise to $O_{5}$ with
\begin{equation}
\Lambda=\frac{M_{U}^{2}}{\lambda_{q}\lambda_{u}\lambda v}.
\end{equation}

As in the case of a scalar exchange there are effects of non-locality,
depending on the value of $M_{U}\approx M_{Q}$. The $\hat{t}$ term in the
$t$-channel fermion propagator, $(\hat{t}-M_{U}^{2})^{-1}$, while negligible
for small $\hat{t}$, suppresses the rate for events with the large invariant
mass. The strength of this suppression can be estimated from $\hat{t}\sim-
\hat{s}/2$ (relevant for a $\sim90^{\circ}$ scattering angle). As discussed in
Section \ref{Production}, $\hat{s}\sim1$ TeV for events near the tail of the
$M_{\text{muons}}$ distribution in Fig.~\ref{fig-O}. Quantitatively, this
effect is illustrated in Fig.~\ref{fig-q2} for $M_{U}=1$ TeV, showing that a
much lower value of the heavy fermion mass would lead to a too strong
depletion of the tail in the invariant mass distribution of the muons. To
comply with $\Lambda\approx100~$GeV, $M_{U}\gtrsim1~$TeV requires therefore
$\lambda_{q}\lambda_{u}\lambda\gtrsim60$, i.e. a collection of pretty strong couplings.

\smallskip

The Lagrangian (\ref{LF}) generates also several 4-quark interactions at one
loop, potentially relevant for the dijet physics. The typical scale of these
operators will be $\sim4\pi\Lambda\approx1.3~$TeV$,$ which, as previously
discussed, should be compatible with current limits.

\subsection{Problems for a more complete theory}

The formulation of a full proper extension of the SM that incorporates in a
consistent way both the hidden and the mediation sectors, in one of the two
versions above, would obviously be premature at this stage and, as such, is
outside the scope of this work. Here we briefly mention some of the problem
one would have to face. They are of at least three kinds:

\begin{enumerate}
\item Flavor problems;

\item Naturalness problems associated with the lightness of $\phi$ (and the
other hidden particles);

\item In the scalar-exchange case, a naturalness problem associated with the
need of a vanishingly small vacuum expectation value of $H$.
\end{enumerate}

The flavor problems are associated with the presence of the effective 4-quark
interactions already discussed in connection with possible dijet signals. Even
in the 2/3-charge sector the limits on the effective scale of these
interactions, if not properly aligned in flavor space, are severe. One at
least technically acceptable way out may consist in assuming that all the
Yukawa interactions in $L_{H}$, Eq. (\ref{LH}), or $L_{F}$, Eq. (\ref{LF}),
\textit{and} in the SM be flavor diagonal \textit{except }for the usual Yukawa
coupling of the \textit{down} quarks, which would therefore be the only source
of flavor breaking.

The issue of the naturally small masses of the hidden sector scalars has a lot
to do with their nature and their possible internal structure, about which
little is known. Nevertheless it may be in any case nontrivial to accommodate
in a natural picture their large Yukawa couplings as in (\ref{LH}) or in
(\ref{LF}).

Finally, from (\ref{LH}) a non vanishing vacuum expectation value of $H$ gives
rise to contribution to the up quark mass, $\delta m_{u}=\lambda
_{q}\left\langle H\right\rangle $, which needs to be kept under control. In
turn, since $\lambda_{q}$ cannot be too small, this is a severe limit on the
mixing squared mass term $m^{2}H^{\dagger}H_{s}$ in the scalar potential. Seen
another way, the operator $O_{5}$ leads to a radiative contribution to the up
quark mass from a $\phi$ loop. This radiative correction could be removed,
without affecting the multi muon signal, by replacing $|\phi|^{2}%
\rightarrow\phi^{2}$ in $O_{5}$ and making proper adjustments in the
deconstructed versions.

\begin{figure}[t]
\begin{center}
\includegraphics[width=\textwidth] {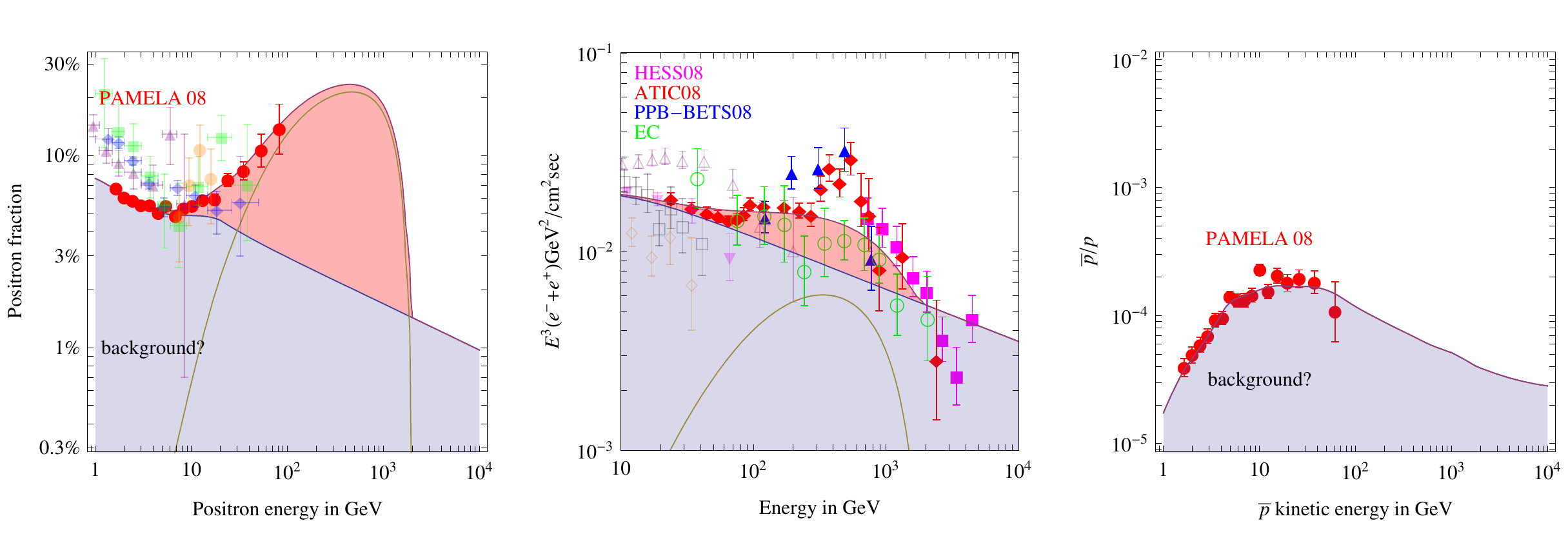}
\end{center}
\caption{\emph{The cosmic ray data about the positron fraction $e^{+}/(e^{+}+e^{-})$ (left), $e^{+}%
+e^{-}$ (middle), anti-proton fraction
$\bar p/p$ (right)  compared to the expected astrophysical backgrounds (lower shaded curve)
plus the Dark Matter excess (upper red curve).}}%
\label{figDM}%
\end{figure}

\section{Dark matter connection?}

Let us assume that the proposed interpretation of the CDF events is correct:
a hidden sector with particles of mass
of order 10 GeV, that decay into $\tau$'s and
coupled to the Standard Model trough new states $Q,U$ or $H$ with
mass in the TeV region. Here we suppose that this TeV sector is augmented by a
stable neutral state $\chi$ that is the dark matter. To connect to the
multi-muon data, we also assume that $\chi$ is also strongly coupled to the
hidden sector by the Yukawa interactions $\chi\chi(\phi,\phi_{1},\phi_{2})$
for fermionic $\chi$, or the quartic couplings $\chi^{\dagger}\chi(\phi^{\ast
}\phi,\phi_{1}^{\ast}\phi_{1},\phi_{2}^{\ast}\phi_{2})$ for scalar $\chi$, so
that the dominant dark matter annihilation channels are
\begin{equation}
\chi\chi\rightarrow\phi^{\ast}\phi,\phi_{1}^{\ast}\phi_{1},\phi_{2}^{\ast}%
\phi_{2}\rightarrow(\bar{\tau}\tau)^{8},(\bar{\tau}\tau)^{4},(\bar{\tau}%
\tau)^{2}.
\end{equation}
If this annihilation cross section is large enough, the $e^{+}$ from $\tau$
decay are able to explain the excess positron cosmic ray signal reported by
PAMELA~\cite{PAMELA}, for $m_{\chi}>(2,1,0.5)$ TeV, for dominant annihilations
to $\phi,\phi_{1},\phi_{2}$ respectively. The predicted spectrum for the
positron excess is shown in Fig.~\ref{figDM}, that applies equally well to
annihilation via $\phi$ or $\phi_{1}$ or $\phi_{2}$ for $m_{\chi}=(4,2,1)$
TeV, respectively. Since the kinematics is such that the $\tau$ are produced
non-relativistically in the rest frame of the decaying hidden sector scalar,
the positron spectrum is similar in the three cases. A smoother spectrum
arises if instead DM has more than one annihilation channel with comparable
branching ratios.
(See~\cite{BCST} and references therein for a discussion of the expected
astrophysical backgrounds and uncertainties).

The connection with the CDF multi-muon data is that the signals involve $\tau$
decays. As a consequence there is no associated $\bar p$ excess, compatibly
with PAMELA observations~\cite{PAMELA}.

Future experiments measuring the positron excess at energies beyond 100 GeV
should find that the excess persists smoothly to higher energies, with a
turnover behavior, shown in Fig.~\ref{figDM}, that is not as sharp as when
annihilations occur directly to electrons or muons.

\medskip

To obtain the PAMELA positron excess, the $\chi\chi$ annihilation cross
section must be larger than that required for a thermal freeze-out abundance
of $\chi$ by a boost factor
\begin{equation}
B \approx10^{3} \frac{m_{\chi}}{\mbox{TeV}}, \label{eq:B}%
\end{equation}
independent of whether the annihilation yields 2,4,or 8 $\bar{\tau}\tau$
pairs. A contribution to this boost factor arises from a Sommerfeld
enhancement factor, from a ladder of $\phi$ exchanges between the initial
annihilating $\chi$ states, and is significant since $m_{\phi}\ll m_{\chi}$.
If $\chi$ is a fermion the Yukawa coupling $\chi\chi\phi$ leads to this
ladder, while for $\chi$ a scalar a new trilinear scalar coupling must be
added, $\chi\chi\phi$. For the case of the Yukawa coupling, $y\, \bar{\chi} (S
+ i\gamma_{5} A) \chi$, the desired thermal freeze-out abundance results if
$y^{2} \approx m_{\chi}/1.5 \mbox{TeV}$. For $m_{\chi}$ in the range of 1 to
10 TeV the Sommerfeld enhancement factor is approximately given by
\begin{equation}
R \approx200 \, \frac{m_{\chi}}{\mbox{TeV}}. \label{eq:R}%
\end{equation}
Comparing with (\ref{eq:B}), this is only a factor 5 below the required boost
factor. Hence if local clumpiness of the halo provide the extra $\approx5$
enhancement, our theory is able to yield the PAMELA positron excess even with
dark matter produced thermally. Such a local clumpiness is helpful in reducing
tensions with limits on gamma~\cite{BCST} and neutrino fluxes.

In Fig.~\ref{figDM} we fixed the DM mass $m_{\chi}$ trying to reproduce the
peak suggested by the ATIC~\cite{ATIC-2} observations of the $e^{+}+e^{-}$
spectrum below 1 TeV. However, the spectral feature produced via $\tau$ decays
(as suggested by the CDF anomaly) is less pronounced than that arising from DM
annihilations to electrons or muons, so that higher statistics data will test
the hypothesis of annihilations via $\tau$ pairs.
\footnote{{\bf Note Added}. The  FERMI experiment~\cite{FERMI} did not confirm the ATIC peak and
observed an $e^++e^-$ spectrum compatible with the model prediction in figure~\ref{figDM}b.}

\section{Conclusions}

We explored the possibilities for a new-physics interpretation of the CDF
multi-muon events, adopting the phenomenological conjecture of
\cite{Giro-small} which links the muons to the pair production of two hidden
sector scalars, each decaying into 8 $\tau$ leptons. Within this conjecture we
addressed the major puzzle left unexplained so far: a plausible production
mechanism which could explain the total muon invariant mass spectrum. The
hardness of this spectrum suggests to explore higher-dimension local operators
connecting two quarks or gluons to a pair of $\phi$'s, and our systematic
search revealed that one such operator, namely the dimension-5 $O_{5}=\bar
{q}q|\phi|^{2}/\Lambda$, reproduces the measured distribution for
$\Lambda\simeq100$ GeV.

\smallskip

Encouraged by this discovery, and by the ability to reproduce several other
experimental distributions, we proceeded to discuss the next crucial question:
is it at all conceivable that such a low scale can be generated in a more
complete theory without producing any other unseen effect? We presented the
two minimal ways of solving this problem, either by exchanging an $s$-channel
new heavy scalar, or by exchanging a $t$-channel new heavy quark. In both
cases 4-quark operators are unavoidably generated along with $O_{5}$, with a
scale of the order a TeV. While the comparison with the existing Tevatron
limits from dijet studies is nontrivial, our preliminary analysis shows that
in both cases no obvious contradiction exists. Furthermore we found that the
effects due to momentum dependence of the exchanged particle propagators may
not disturb the agreement of the effective-operator limit with the total
invariant mass distribution.

\medskip We think that all this offers a definite and consistent framework
which can be subject to stringent tests by further comparison with the data.
In fact, since the production mechanism is now largely fixed, it should be
possible for the experimentalists to pronounce a final word on this model. We
propose possible tests involving muons only; more tests involving hadronic
tracks can be imagined.

\medskip

If we assume that the proposed interpretation of the CDF events is correct,
further work in many different directions can and must be done. Here we have
not resisted from drawing a possible connection between the CDF data and the
putative Dark Matter signals seen in PAMELA and/or ATIC. Assuming that Dark
Matter is a hidden sector particle that annihilates into the $\phi$-scalars
decaying into $\tau$ leptons, we explored its indirect signals: one can
reproduce the PAMELA $e^{+}$ excess and, at higher energies, a feature in the
$e^{+}+e^{-}$ spectrum milder than the one hinted at by the ATIC data.

\subsection*{Acknowledgements}

We are especially grateful to Paolo Giromini for many discussions and
invaluable clarifications about the multi-muon CDF data. We thank Luciano
Ristori and Michelangelo Mangano for discussions about the multi-muon events,
and Gennaro Corcella for discussions about Monte-Carlo generators. R.B. and
V.R. are partially supported by the EU under RTN contract MRTN-CT-2004-503369
and by MIUR under the contract PRIN-2006022501. The work of LH is supported by
the U.S. Department of Energy under contract no. DE-AC02- 05CH11231 and NSF
grant PHY-04-57315.

\appendix

\section{Explicit realizations for $O_{6G}$ and $O_{6F}$}

\subsection{Gluon operator}

To \textit{deconstruct} $O_{6G}$ one needs at least one heavy quark singlet
$F$, of mass $M_{F}$ and unspecified charge, and the mediation coupling
$\lambda\phi\bar{F}F$. Integrating it out at one loop gives rise to
\begin{equation}
{L}_{\mathrm{eff}}\approx\frac{\lambda^{2}g_{s}^{2}}{16\pi^{2}M_{F}^{2}}%
O_{6G},
\end{equation}
so that the threshold effect in Fig.~\ref{fig-O}b could be obtained with
$M_{F}\sim(\lambda/4\pi)\times200$ GeV.

\subsection{Fermion operator}

As in the case of $O_{5}$, there are two ways to generate $O_{6F}$, either by
fermion or by heavy vector exchange. Concentrating on the vector-exchange
case, the relevant gauge Lagrangian for a $Z^{\prime}$ of mass $M$ could be
\begin{equation}
{L}=g_{q}Z_{\mu}^{\prime}\sum_{i}\bar{q}_{i}\gamma_{\mu}q_{i}+g_{\phi}Z_{\mu
}^{\prime}(\phi^{+}\overleftrightarrow{\partial}_{\mu}\phi)
\end{equation}
where $i$ runs over all quarks. This can account for $O_{6F}$, with the needed
effective scale to explain the tail of Fig.\thinspace\ref{fig-O}b, provided
\begin{equation}
M\approx(g_{\phi}/4\pi)^{1/2}\sqrt{g_{q}}\,700~\text{GeV}.
\end{equation}
At the same time, the $Z^{\prime}$-width is dominated by decays into $\phi
\phi^{\ast}:$%
\begin{equation}
\Gamma\sim\frac{g_{\phi}^{2}}{48\pi}M.
\end{equation}
We believe that this leads to an acceptable situation, also in view of the
dijet limits from CDF, provided $g_{\phi}\rightarrow4\pi$. A model like this
might be the easiest to incorporate in a complete extension of the SM.

\end{document}